# Synthesis of copper and silver nanoparticles by molecular beam method

Yu.A. Kurapov[1], S.E. Litvin[1*], S.M. Romanenko[1], G.G. Didikin[1],
N.N. Belyavina[2]

[1]E. O. Paton Electric Welding Institute of the National Academy of Science of Ukraine,
11 Kazymyr Malevych St., Kyiv, 03150, Ukraine
[2]Department of Physics, Taras Shevchenko University,
4 Glushkov Ave., Kyiv, 03022, Ukraine

*Corresponding author:
Dr. Stanislav Litvin    e-mail:    litvin@paton.kiev.ua

**Abstract**

The paper presents the results of investigation of the structure of porous condensates of Ag-NaCl and Cu-NaCl composition; chemical and phase compositions and dimensions of nanoparticles, produced from the vapour phase by EBPVD method. Silver and copper nanoparticles in a porous matrix, when removed from vacuum, oxidize in air to lower oxides, and have considerable sorption capacity relative to oxygen and moisture. Heating in air is accompanied by lowering of porous condensate mass, primarily, due to desorption of physically sorbed moisture, as well as afteroxidation to higher oxides due to physically adsorbed oxygen. At copper concentrations up to 10 at.%, sorption capacity of the condensate is greatly enhanced, that is attributable to presence of very fine nanoparticles. Increase of copper concentration in the condensate is accompanied by lowering of sorption capacity of nanoparticles with respect to oxygen, and, therefore, also change of phase composition. In addition to concentration change, phase composition of nanoparticles can be also controlled by heat treatment of the initial condensate produced at low condensation temperatures. Silver and copper nanoparticles can be converted into stable colloidal systems.

## 1. Introduction

Among the currently produced nanomaterials for medical applications, preparations with silver and copper nanoparticles attract special attention. It is known that silver has stronger antimicrobial properties than does penicillin, biomycin and a number of other antibiotics, having a devastating effect on antibiotic resistant bacterial strains [1-3]. Moreover, silver aquasol has a synergic effect, extending the term of action of some antibiotics many times [1, 4, 5]. The antiseptic effect is achieved at application of a smaller quantity of antibiotic. It is also shown that nanosilver stimulates the immune system, stabilizes the metabolism in the living body and disinfects about 100 species of dangerous bacteria, viruses and fungi [6, 7], while the antibacterial spectrum of any antibiotic extends to 5-10 strains of microorganisms.

Here, oxides of these metals are better antibacterial agents, degree of metal oxidation playing a significant role [7-9]. It is found [10-12] that in addition to silver, nanoparticles of copper and copper oxide also exert a pronounced antibacterial effect on gram-positive and gram-negative bacteria. Copper itself is a vital (i.e. essential) element for living organisms. In particular, it is a cofactor of more than 20 enzyme systems that provide both the mammalian and human multiple processes of vital activity, and the body's resistance to bacterial infections and other invasions. Although copper has less pronounced antiseptic properties than silver, it can significantly enhance the effect of silver preparations [6, 13].

Nanoparticles, produced practically by all the methods, are in a metastable non-equilibrium state. This circumstance complicates their study and use in nanotechnology to create stably operating devices. On the other hand, the non-equilibrium of the system enables conducting new, unusual and hardly predictable chemical transformations. For small particles, specific size effects are most strongly manifested, where irregular dependencies of properties on particle size prevail.

The degree of activity of the nanoparticle, depending on its size, is due to its changing properties when interacting with the adsorbed reagent [14]. The main adsorbed reagent for metal nanoparticles is oxygen.

Further development and creation of medicines based on silver and copper nanoparticles require selecting a method for synthesis of nanoparticles that would ensure production of various nanomaterial compositions based on these metals with a specified content of the necessary structural components.

In most cases, the physical methods for synthesizing nanoparticles are based on physical processes of evaporation or atomizing of substances by highly concentrated energy sources followed by deposition of the vapor phase (atomic-molecular beams) in vacuum, or atmosphere of inert or active gases. In this respect, the method of electron beam evaporation and deposition of materials in vacuum opens up wide possibilities [15, 16]. The idea of using this approach is the possibility of simultaneous evaporation and deposition of mixed molecular flows of two substances onto a substrate. By lowering the temperature of the substrate, it is possible to ensure the limitation of diffusion mobility of atoms in the solid state and, thereby, to create conditions for the formation of a composite structure with the necessary element ratios and their dispersion. Water-soluble substances, for example sodium chloride [17-19], are used as the main component (matrix), that gives an advantage in further use of these substances for the preparation of colloidal solutions.

This paper presents the results of studying the processes of physical synthesis of silver and copper nanoparticles by molecular beams, their thermal stability at interaction with air oxygen, and the method of conservation of deposited particles in a salt matrix designed to store particles in an un-aggregated state for further use in medicine.

**2. Experimental procedure**

A metal ingot and compacted NaCl cylinder were placed into cylindrical water–cooled crucibles of 50 mm diameter, located side-by-side. A vacuum of $1.3 – 2.6 \times 10^{-2}$ Pa was created in the reaction chamber. Both the materials were heated in the crucibles by electron beam gun up to complete melting. This resulted in formation of a mixed vapor flow of metal and NaCl particles, which condensed on a water-cooled substrate at the temperature of 40-50 °C. After bleeding air and complete unsealing of the chamber, the condensate was scraped off the substrate, and the deposited nanoparticles were studied both in dry condensate and in a colloidal solution after their dissolution.

X-ray diffraction (XRD) data for carrying out qualitative phase analysis (including lattice parameters refinement and crystal structure refinement) were collected with DRON-3 automatic diffractometer (CuK$\alpha$ radiation). The diffraction patterns have been obtained in a discrete mode under the following scanning parameters: observation range $2\theta = (20–120)°$, step scan of 0.05°, and counting time per step at 3 s.

The microstructure of the condensate and its elemental content were examined with both the CamScan (Cambridge CamScan SEM, UK) scanning electron microscope (SEM) and the INCA-200 energy-dispersive X-ray imager (Oxford Inca Energy 200 EDS system, UK) and the transmission electron microscopy (TEM) method using H-800 microscope (Hitachi, Japan) at an accelerating voltage of 100 kV. The air oxidation of metallic nanoparticles placed in the salt matrix was studied using a thermogravimetric analyzer TGA-7 (Perkin Elmer, USA) (heating/cooling at 10 °C / min to 650 °C). The size of the nanoparticles in the colloidal system was studied by dynamic light scattering (DLS) [20, 21] using the Zeta Sizer-3000 laser correlation spectrometer (Malvern, UK).

# 3. Investigation results and discussion

## 3.1. Deposition of Cu-NaCl and Ag-NaCl condensates on the substrate

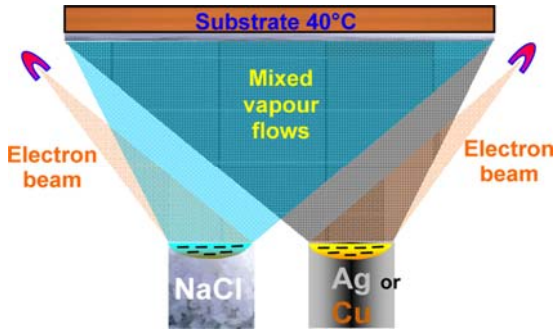

Fig. 1. Process schematic of mixing the vapour flows of salt (NaCl) and metal (Ag or Cu) in vacuum and their deposition on a stationary water-cooled substrate with temperature $T_S$.

Condensate synthesis was performed in the work by vacuum deposition of mixed molecular flows of metal and salt particles on the substrate by the technological schematic shown in Fig. 1.

Realization of this schematic allows producing on a substrate, heated up to $< 0.3\ T_m$, the initial condensate of salt with embedded metal nanoparticles and microstructure with open micro- and nanosized porosity [22, 23].

Studying the transverse cleavage of porous Cu-NaCl condensate by EDS method showed a uniform distribution of elements (copper, sodium, chlorine and oxygen) across the condensate thickness (Fig. 2), that is indicative of maintaining a constant ratio of evaporation rates for individual components of the vapour flow during the entire technological process.

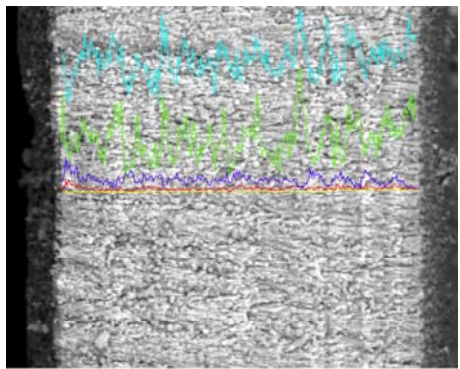

(a)

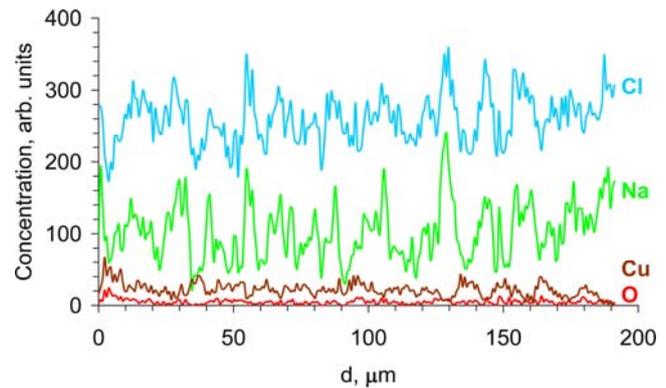

(b)

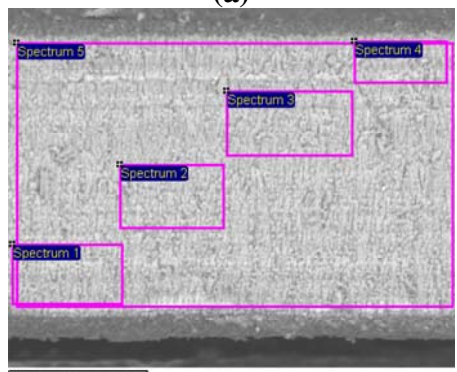

(c)

| Spectrum | Cu | O | Na | Cl |
|---|---|---|---|---|
| | wt % | | | |
| Spectrum 1 | 21.0 | 4.0 | 27.0 | 48.0 |
| Spectrum 2 | 20.0 | 4.0 | 27.5 | 48.5 |
| Spectrum 3 | 19.2 | 4.0 | 27.6 | 49.2 |
| Spectrum 4 | 14.6 | 4.1 | 31.1 | 50.2 |
| Spectrum 5 | 19.8 | 4.0 | 27.3 | 49.0 |

(d)

Fig. 2. Transverse cleavage microstructure (a), element distribution across the thickness (b) and local element analysis (c, d) of Cu-NaCl condensates produced on a water-cooled copper substrate with 40–50 °C temperature. In (a) substrate is on the left, in (c) substrate is below.

## 3.2. Deposited Cu-NaCl condensate

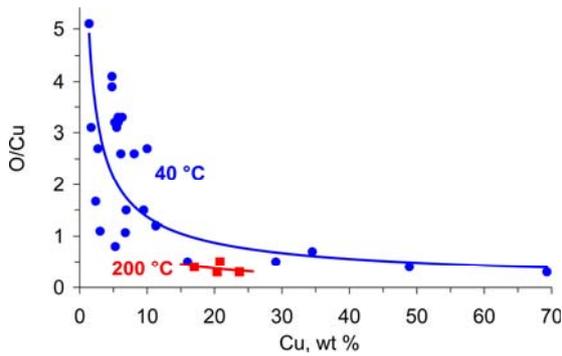

Fig. 3. Dependence of O/Cu index on copper content in the condensates of Cu-NaCl composition, produced at different temperatures $T_S$: 40 °C and 200 °C.

Results of EDS analysis are indicative of presence in Cu-NaCl samples of a certain quantity of oxygen, adsorbed by metal nanoparticles (Figs. 2, 3). Atomic ratio of oxygen to copper content (O/Cu index) in the condensate decreases both at increase of copper content, as at increase of substrate temperature (Fig. 3).

A set of data of EDS analysis, scanning electron microscopy of thin chips and XRD studies indicate the presence of a fine nanoscale substance in the initial Cu-NaCl condensates (the size of individual particles is about 20 nm), whose phase composition corresponds to a mixture of NaCl and $Cu_2O$ (Figs. 4, 5, Table 1). It should be noted that NaCl lattice period for all the samples studied within the error, corresponds to the lattice period of pure common salt a = 0.5640(2) nm.

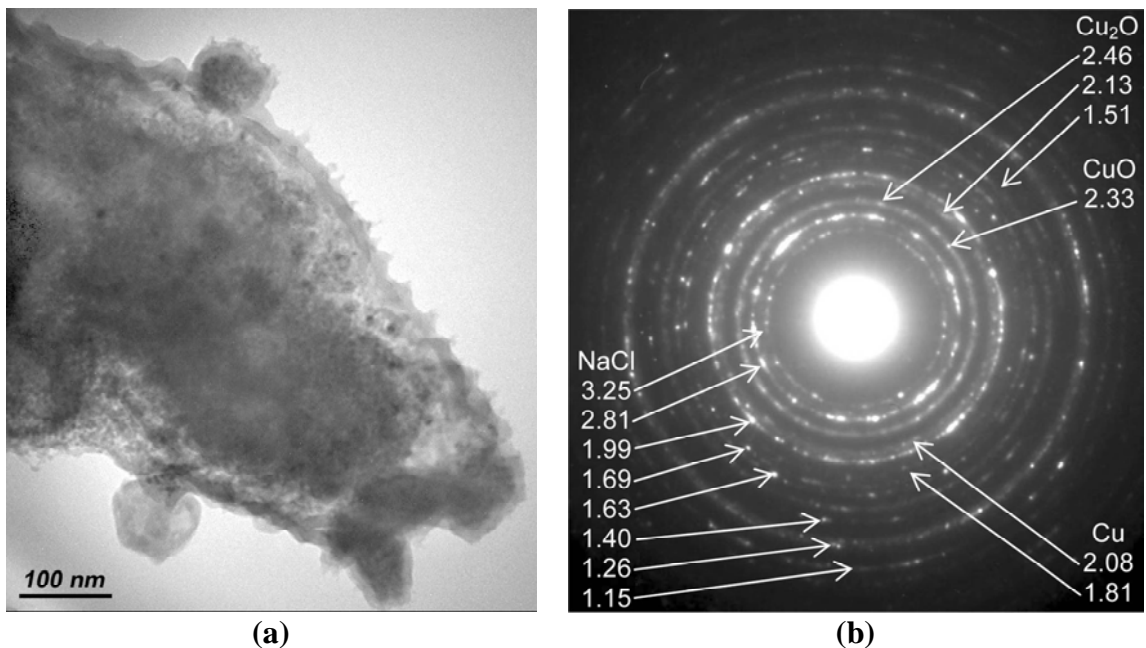

(a) (b)

Fig. 4. Microstructure (a) and electron diffraction pattern (b) of initial Cu-NaCl condensates.

Analysis of the above data is indicative of the fact that at breaking the vacuum in the chamber the phase composition of initial condensates is mainly determined by the degree of aggregation of nanoparticles, both in the mixed vapour flow and at their condensation on the substrate (Table 1). So, at a small content of copper (up to ≈ 20 wt %), and, hence at low probability of nanoparticle aggregation, the condensate contains $Cu_2O$ oxide. At average copper content (≈ 20 – 30 wt %) the condensate predominantly contains a mixture of $Cu_2O$ and CuO phases, alongside NaCl. And, finally, because of the high probability of nanoparticle aggregation at a high content of copper (≈ 30 – 50 wt %) the fraction of coarser particles increases, and CuO and Cu phases co-exist with $Cu_2O$ phase.

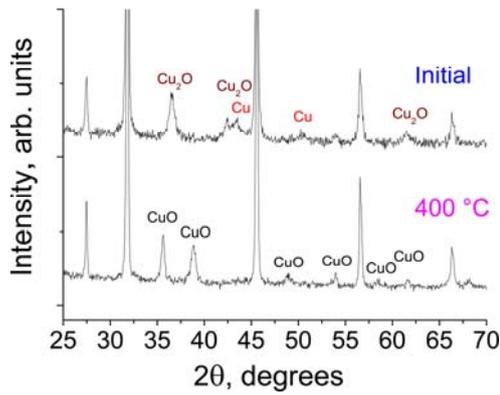
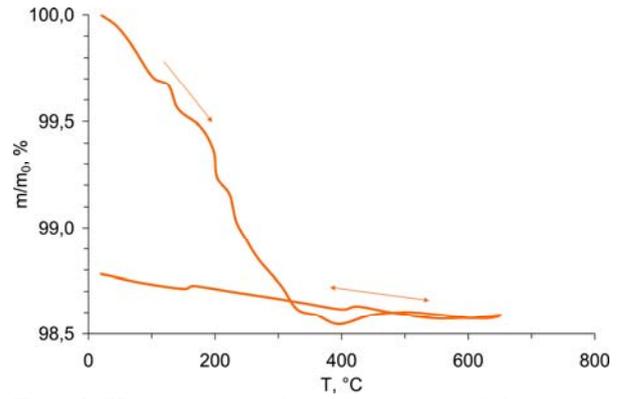

Fig. 5. Diffractograms of Cu-NaCl condensate in the initial condition and after annealing at 400 °C on air (phase designation: $Cu_2O$, Cu, CuO, balance is NaCl).

Fig. 6. Kinetics of relative change of the mass of crushed Cu-NaCl condensate (13.7 wt % Cu) at two cycles of heating and cooling in air.

Table 1

Phase composition of initial Cu-NaCl condensates

| Cu, wt % | Phase composition |
|---|---|
| 13.7 | NaCl (86) + $Cu_2O$ (10) + CuO (4) |
| 23.0 | NaCl (77) + $Cu_2O$ (23) |
| 26.0 | NaCl (76) + $Cu_2O$ (24) |
| 27.5 | NaCl (72) + $Cu_2O$ (28) |
| 30.0 | NaCl (70) + $Cu_2O$ (27) + CuO (3) |
| 32.6 | NaCl (67) + $Cu_2O$ (23) + CuO (5) + Cu (5) |
| 33.1 | NaCl (67) + $Cu_2O$ (23) + CuO (4) + Cu (6) |
| 47.1 | NaCl (53) + $Cu_2O$ (32) + CuO (8) + Cu (7) |

$T_s$ = 30–50°C, condensate thickness of 125–200 μm. Phase content, wt %, is given in brackets

Table 2

Phase content of Cu-NaCl condensate (13.7 wt % Cu) after its heat treatment

| Heat treatment | Phase composition |
|---|---|
| Initial | NaCl (86) + $Cu_2O$ (10) + CuO (4) |
| 300 °C, 1h | NaCl (85) + CuO (15) |
| 400 °C, 1h | NaCl (86) + CuO (15) |

Phase content, wt %, is given in brackets.

Therefore, at loss of vacuum in the chamber physical adsorption of oxygen by open surface of active nanoparticles of copper, embedded into the micro- and nanoscale pores of the salt matrix, takes place with formation of $Cu_2O$ and CuO oxides. Coarsening of particles, occurring at increase of copper concentration in the condensate, leads to individual metal particles consisting of a copper nucleus coated by an oxide shell.

Heat treatment of the initial components significantly affects the phase composition (Table 2).

It is shown that increase of the temperature of initial condensate treatment promotes acceleration of diffusion processes in the material, leading to afteroxidation of copper particles (from $Cu_2O$ to CuO). The oxidation process was studied in detail by the method of gravimetric analysis by the kinetics of relative change of mass of porous Cu-NaCl condensate (13.7 wt % Cu) at heating up to 650 °C and cooling in air (Fig. 6).

As a result it was shown that with increase of temperature (at heating at the rate of 10 °C/min) lowering of porous condensate mass occurs right up to 400 °C temperature, due to removal from it of water $H_2O$, intercrystalline moisture and OH hydroxyl groups. Further increase of temperature (above 400 °C), is accompanied by a slight increase of porous condensate mass due to afteroxidation of $Cu_2O$ to CuO by air oxygen (Fig. 2). At heating to about 650 °C, just CuO is present in the condensate, so that further heating cycles are not accompanied by any changes (Fig. 6). It is obvious that annealing in air for one hour at the temperature of 300 °C, ensures running of the process of $Cu_2O$ afteroxidation with formation of CuO oxide in the condensate (Table 2).

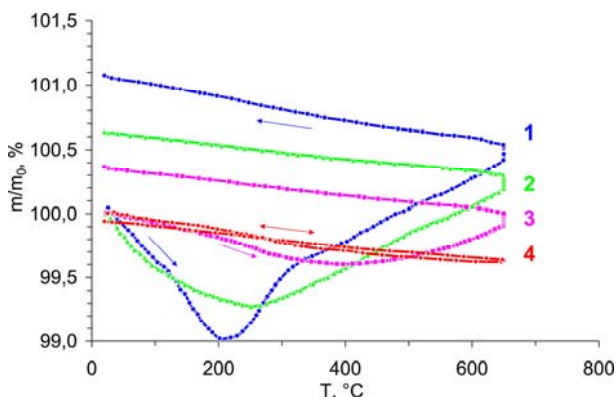

Fig. 7. Kinetics of oxidation of Cu-NaCl condensates (22.1 wt % Cu) at two cycles of heating and cooling in air, depending on annealing temperature $T_{an}$: 1 – initial, 2 – 200 °C, 10 min., 3 – 400 °C, 13 min., 4 – 650 °C, 5 min.

With increase of copper content (22.1 wt %) in the condensate, the fraction of adsorbed moist air decreases due to increase of the size of nanoparticles and reduction of their activity (Fig. 7, curve 1).

Due to a greater content of $Cu_2O$ in the initial sample with 22.1 wt % Cu (Table 1), its afteroxidation requires more oxygen. Therefore, after 200 °C, a considerable increase of the mass of condensate with 22.1 wt % Cu is observed right up to 650 °C, (Fig. 7, curve 1). Partial afteroxidation with formation of a mixture of $Cu_2O$ and CuO occurs already during annealing (with temperature rise). And, therefore, less oxygen is required for afteroxidation of remaining $Cu_2O$ (Fig. 7, curves 2, 3) and sample growth decreases. And, finally, annealing at the temperature of 650 °C for 5 minutes leads to complete afteroxidation of $Cu_2O$ phase to CuO (Fig. 7, curve 4), and there is no growth.

Thus, a comprehensive study of Cu-NaCl condensates showed that after breaking vacuum in the chamber the condensate deposited on the substrate, in addition to NaCl, also contains a mixture of $Cu_2O$ and/or CuO oxides, the ratio of which is determined both by concentration of the initial mixture and temperature modes of synthesis and heat treatment.

### 3.3. Deposited Ag-NaCl condensate

According to the data of X-Ray phase and TEM analyses, all the synthesized and heat-treated condensates are two-phase and contain a mixture of NaCl and silver (Figs. 8, 9, Table 3). TEM study of thin chips of porous Ag-NaCl condensate revealed presence of a nanoscale substance of 10 nm size (Fig. 9).

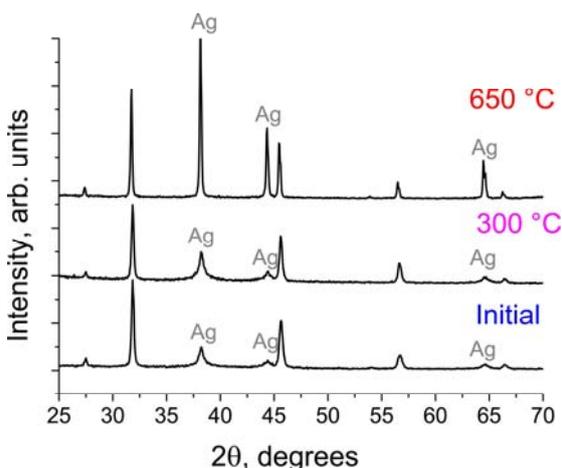

Fig. 8. Diffractograms of Ag-NaCl condensate in the initial condition and after annealing at $T_{an}$ = 300 and 650 °C (phase designation: Ag, balance is NaCl).

It should be noted that according to XRD data, the silver lattice periods at these transformations are constant within the error, while the period of NaCl lattice increases significantly at annealing (Table 3, Fig. 10).

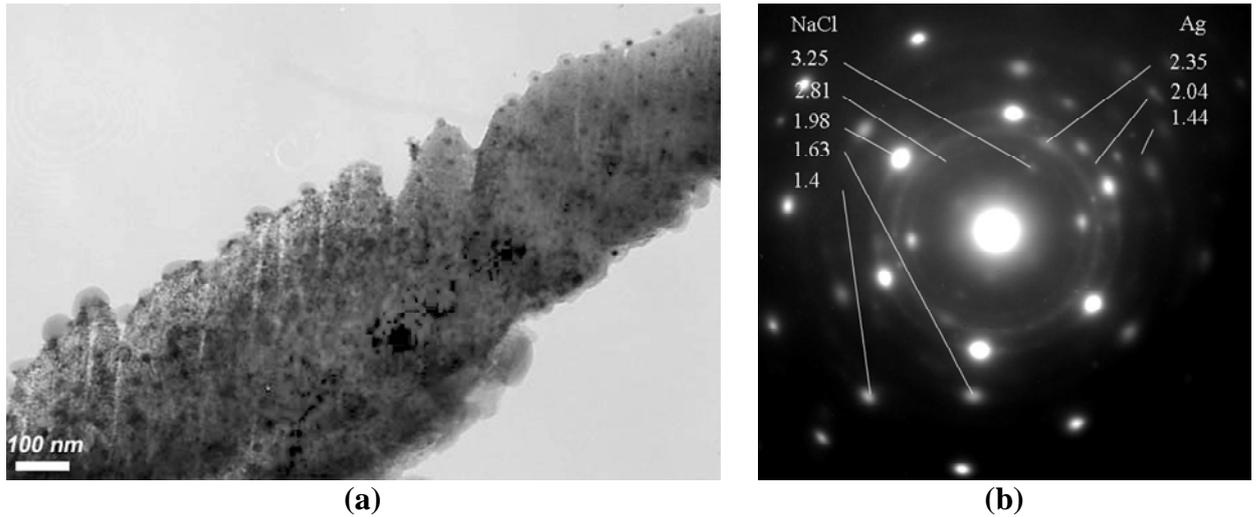

| (a) | (b) |

Fig. 9. Microstructure (a) and electron diffraction pattern (b) of initial Ag-NaCl condensates.

To clarify the nature of variation of the lattice period (Fig. 10), the crystalline lattice of sodium chloride was determined more precisely for Ag-NaCl condensate in the initial condition and after its annealing at 650 °C (Table 4).

Table 3

Characteristics of the phases according to XRD analysis of Ag-NaCl

| Sample | Phase composition | Ag | | NaCl |
|---|---|---|---|---|
| | | Lattice parameter, $a$, nm | Grain sizes, $D$, nm | Lattice parameter, $a$, nm |
| Varying the substrate temperature, $T_s$ | | | | |
| Initial | Ag (45) + NaCl (55) | 0,4088(2) | 6(1) | 0,5636(2) |
| 200 °C | Ag (48) + NaCl (52) | 0,4084(2) | 5(1) | 0,5630(2) |
| 400 °C | Ag (45) + NaCl (55) | 0,40859(3) | 14(1) | 0,56368(3) |
| Pounded powder mixture | | | | |
| Pounded | Ag (50) + NaCl (50) | 0,40878(7) | 8(1) | 0,56368(8) |
| Varying the annealing temperature, $T_a$ | | | | |
| Initial | Ag (45) + NaCl (55) | 0,4085(2) | 9(1) | 0,5628(4) |
| 100 °C 1h | Ag (47) + NaCl (53) | 0,4086(2) | 9(1) | 0,5633(1) |
| 200 °C 1h | Ag (47) + NaCl (53) | 0,4088(1) | 10(1) | 0,5636(1) |
| 300 °C 1h | Ag (47) + NaCl (53) | 0,40859(3) | 15(2) | 0,56378(3) |
| 400 °C 1h | Ag (45) + NaCl (55) | 0,40861(1) | 25(2) | 0,56391(6) |
| 650 °C 1h | Ag (62) + NaCl (38) | 0,40859(1) | coarse-grained | 0,56412(3) |

Obtained results (Table 4) show that a substitutional solid solution of $Na_{0.98}Ag_{0.02}Cl$ forms on the substrate surface at interaction of molecular flows of salt and silver. During annealing of Ag-NaCl condensate, silver precipitates from this solid solution lattice, and at 650 °C salt is present in the annealed condensate already in its initial condition.

Table 4

Crystal data for NaCl compounds existing in the initial Ag-NaCl composite and after its annealing $T_a$ at 650 °C

| Atom | Site | Site occ. | X | y | z |
|---|---|---|---|---|---|
| Initial Ag-NaCl | | | | | |
| Na | *4a* | 0.98(1) | 0 | 0 | 0 |
| Ag | *4a* | 0.02(1) | 0 | 0 | 0 |
| Cl | *4b* | 1.00(1) | 0.5 | 0.5 | 0.5 |

| | |
|---|---|
| Space group | *Fm3m* (no. 225) |
| Lattice parameter, *a*, nm | 0.5627(2) |
| Independent reflections | 15 |
| Total isotropic *B* factor, nm² | $B = 3.28(2) \cdot 10^{-2}$ |
| Reliability factor | $R_I = 0.053$ |

| Atom | Site | Site occ. | X | y | z |
|---|---|---|---|---|---|
| Annealing at 650 °C | | | | | |
| Na | *4a* | 1.00(1) | 0 | 0 | 0 |
| Cl | *4b* | 1.00(1) | 0.5 | 0.5 | 0.5 |

| | |
|---|---|
| Space group | *Fm3m* (no. 225) |
| Lattice parameters, *a*, nm | 0.56391(3) |
| Independent reflections | 15 |
| Total isotropic *B* factor, nm² | $B = 1.49(1) \cdot 10^{-2}$ |
| Reliability factor | $R_I = 0.043$ |

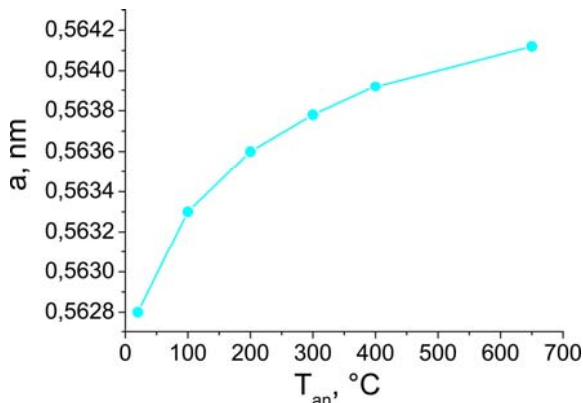

Fig. 10. Dependence of NaCl lattice period *a* on annealing temperature $T_{an}$ of Ag-NaCl condensate.

Average grain sizes of Ag phase were determined from line broadening using the classical Williamson–Hall method. The Williamson-Hall graphs have been plotted as dependencies of scaled broadening of Bragg's reflections

$$b^*(2\theta) = \frac{\beta(2\theta)\cos\theta}{\lambda}$$

on scattering vector $S=(2\sin\theta)/\lambda$ for each sample studied. The average grain size D could be found by extrapolating the $b^*(2\theta)$ dependencies onto $S=0$ axes as $D=1/b^*(2\theta)$ at $\theta=0$ (Table 3, Fig. 11). Thus, according to X-Ray data silver is in the nanostate (about 8 nm grain size) in the initial condensate. During heat treatment the grain size becomes larger, the most significantly with annealing (Fig. 11).

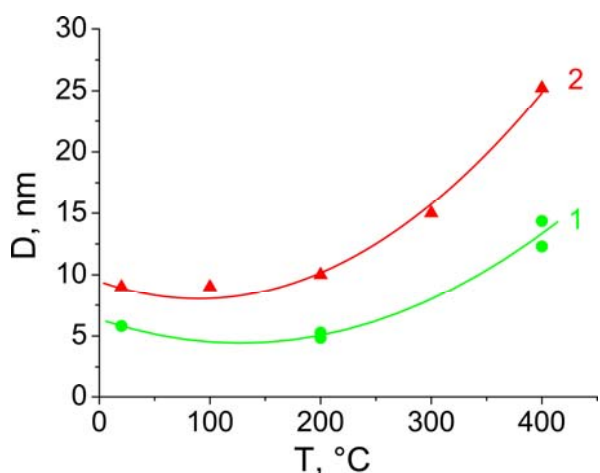

Fig. 11. Silver grain size D in Ag-NaCl condensate, depending on temperature T: 1 – substrate temperature $T_S$; 2 – annealing temperature $T_{an}$.

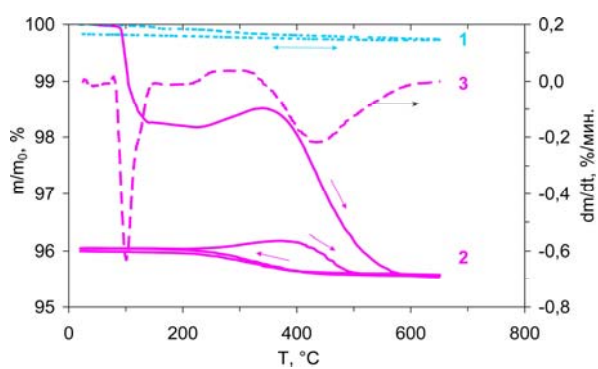

Fig. 12. Kinetics of relative change of mass of NaCl condensate (1) and Ag-NaCl condensate (2) with differential curve (3) at cyclic heating and cooling in air.

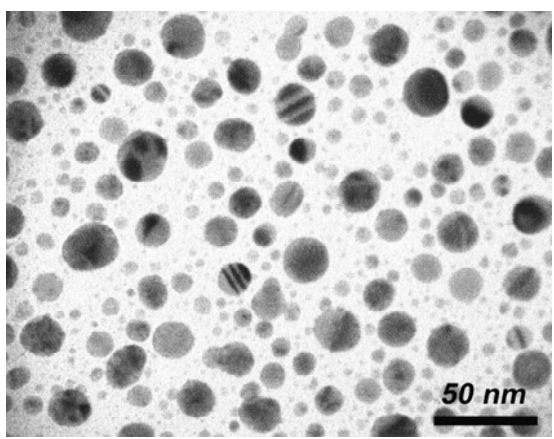

Fig. 13. TEM image of Ag-NaCl condensate annealed at 400 °C (1 hour) and rinsed to remove NaCl for 5 days. Ag twins are visible.

It is known that metallic silver can be covered by a thin layer of $Ag_2O$ oxide in the air atmosphere [4]. However, because of certain limitations for analysis of small quantities of substances, neither X-Ray diffraction method, not TEM found oxygen in Ag-NaCl condensates. Therefore, TGA method was used to study the kinetics of relative change of mass of both porous NaCl condensate, and Ag-NaCl condensate at their heating up to 650 °C and cooling in air. As a result it is shown that the mass of porous NaCl condensate practically does not change during the experiment (Fig. 12, curve 1), i.e. at electron beam evaporation the salt condenses on the substrate at its stoichiometric composition and does not absorb moisture.

Investigations of porous Ag-NaCl condensate performed by TGA method, showed that a slight lowering of its mass occurs in the temperature range of 80 – 120 °C. The 120-650 °C region is characterized by a slight increase of sample mass (maximum at 360 °C) with its subsequent lowering. At subsequent cooling and repeated heating-cooling cycles the condensate mass does not change (Fig. 12, curve 2).

Considering the data of work [4], the possibility of formation of an oxide shell on the surface of silver nanoparticles was assessed visually by the colour of deposited condensate. That is, the condensate, in our opinion, gets its hue from the thinnest film of brown-black crystals of the most thermodynamically stable $Ag_2O$ oxide [24, 25]. Also probable is the presence of $Na_2O$, AgOH and AgCl. Then, change of sample mass in the temperature range of 80 – 120 °C can be associated with removal of adsorbed moisture due to decomposition of metastable AgOH phase into oxide and water. A slight increase of sample mass in the temperature range of 250 – 360 °C, is associated, apparently, with afteroxidation of sodium oxide ($Na_2O \rightarrow Na_2O_2$) and silver oxide ($Ag_2O \rightarrow AgO$). And, finally, lowering of sample mass in the temperature range of 340 – 600 °C (430 °C by differential curve 3) is associated with decomposition of AgO oxide into silver and oxygen. This result is in agreement with the data of Roy [4] on decomposition of AgO oxide at 420 °C (for chemical reactions in aqueous medium values of 200, 250 and 300 °C are given [24, 25]).

Fig. 13 shows TEM microstructure of two-phase Ag-NaCl condensate annealed for 1 hour at 400 °C and rinsed to remove NaCl salt for 5 days. One can see that individual silver particles are of a round shape and 8 to 25 nm size.

Summing up of the results of the conducted comprehensive study leads to the conclusion that Ag-NaCl condensate deposited on the substrate, contains a mixture of metallic silver nanoparticles (of about 8 nm size and, probably, coated by the thinnest $Ag_2O$ shell) and $Na_{0.98}Ag_{0.2}Cl$ solid solution. Heat treatment of the initial condensate leads to coagulation and coarsening of Ag particles, as well as precipitation of finest Ag particles from the lattice of solid solution based on sodium chloride.

### 3.4. Aqueous colloidal solutions of Me+NaCl

It is shown that after addition of water to synthesized Me+NaCl condensates, sodium chloride dissolves, and Me nanoparticles, sticking together into aggregates in the form of bundles or chains (Fig. 14), precipitate during sedimentation.

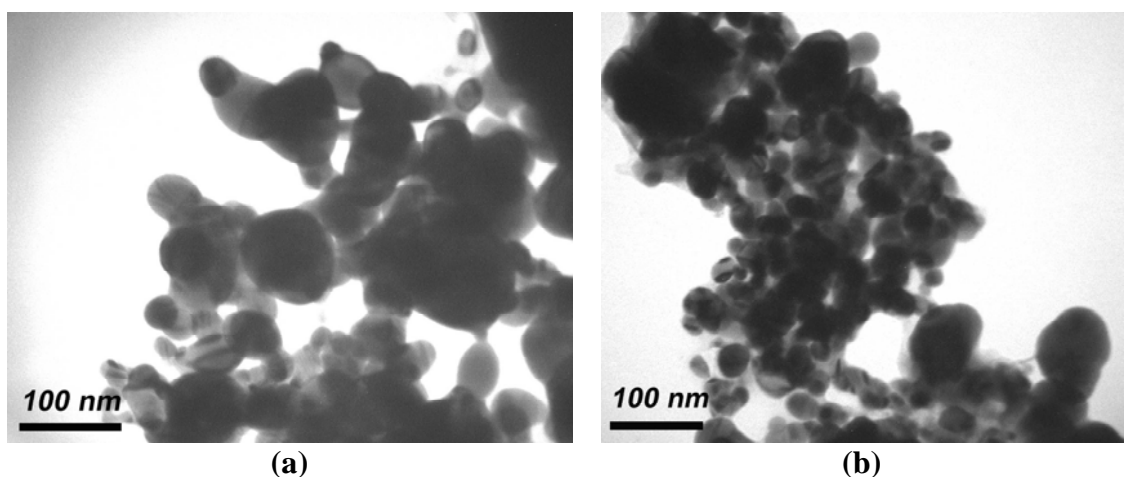

(a)          (b)

Fig. 14. Nanostructure of aggregates of Ag nanoparticles of initial Ag-NaCl condensates (a) and those annealed at 200 °C for 1 hour (b), which were rinsed in distilled water to remove salt.

In order to create adsorption layers, preventing particles coarsening as a result of their sticking together, a stabilizer – a surfactant (SA) was added to the colloid. A 1 vol. % aqueous solution of polyvinylpyrrolidone (PVP) was used as SA for silver particles. No agglomeration of silver particles was observed, and the size of silver nanoparticles suspended in SA aqueous solution (determined by DLS method) (Fig. 15), corresponded to the size of nanoparticles in the initial Ag-NaCl porous condensate (about 10 nm).

For copper oxide nanoparticles formed in Cu-NaCl condensates 1 vol. % solution of polyvynil alcohol (PVA) was used as SA. Quantitative distribution of particles by sizes, obtained by DLS method (Fig. 16), reveals a maximum at 14-16 nm that corresponds to the size of nanoparticles in porous Cu-NaCl condensate. DLS results are confirmed by electron microscopy studies of

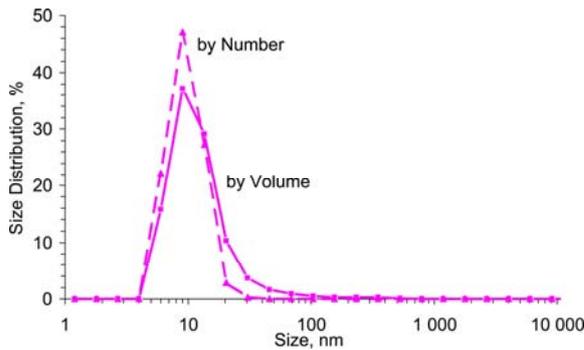 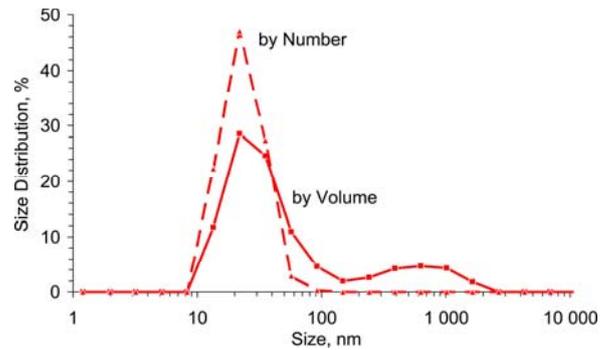

Fig. 15. Distribution of particle size in 1 vol.% aqueous solution of polyvynilpyrollidone and of initial Ag-NaCl condensate.

Fig. 16. Distribution of particle size in 1 vol.% aqueous solution of polyvynil alcohol and initial Cu-NaCl condensate.

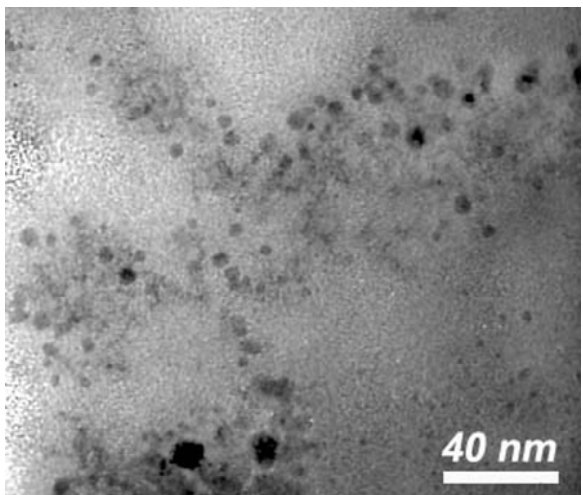 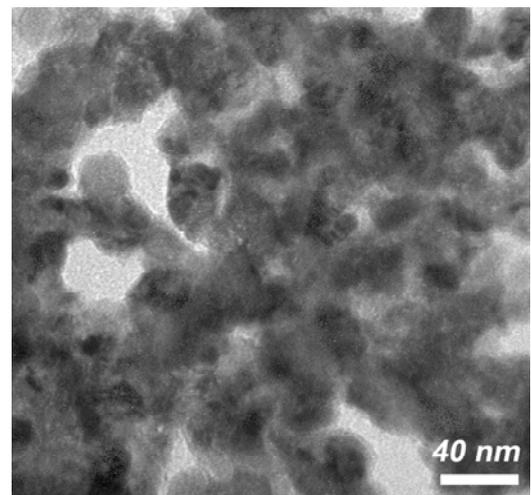

**(a)** **(b)**

Fig. 17. Nanostructure of dried aqueous solution with nanoparticles of silver (a) and copper (b) after rinsing of initial Ag-NaCl (a) and Cu-NaCl (b) condensates to remove salt.

dried solutions (Fig. 17). Thus, using colloidal aqueous solutions with SA additives, it was possible to obtain isolated nanoscale particles of silver and copper oxide (Fig. 17).

**4. Conclusions**

Physical synthesis of silver and copper nanoparticles by the method of condensation of mixed molecular beams of metal and salt in vacuum allowed producing nanostructured material in the form of a dry substance (silver and copper nanoparticles in a water-soluble matrix), which enables preservation, storage and transportation of particles for preparation of the required colloidal solutions.

Low thermodynamic stability of the compounds of silver with oxygen did not allow their direct identification by TEM and XRD methods, although TGA method provides an indication of the processes of transformation in Ag-O system, and, if we assume that silver nanoparticles have a very thin film of silver oxide in the form of $Ag_2O$ on their surface [4], it decomposes at the temperature of 460 °C.

Controlling the activity (size) of nanoparticles by varying copper concentration in the condensate, annealing temperature and duration, allows creating various compositions of nanomaterials, based on copper with preset content of the required structural components.

**List of symbols**

T – temperature
$T_m$ – melting temperature
$T_{an}$ – annealing temperature
$T_S$ – substrate temperature
*D* -  grain size
*a* - lattice parameter (period)
d – condensate thickness
$m/m_0$ - relative change of mass of TGA
*B* - total isotropic factor
$R_I$ - reliability factor
EBPVD - electron-beam physical vapor deposition
XRD - X-ray diffraction
SEM - scanning electron microscope
TEM - transmission electron microscopy
DLS - dynamic light scattering
EDS - energy-dispersive X-ray spectroscopy
TGA - thermogravimetric analyzer
SA – surfactant
PVP – polyvinylpyrrolidone
PVA - polyvynil alcohol